\documentclass[aps,prl,twocolumn,groupedaddress]{revtex4}
\usepackage{graphicx}
\begin{document}

\title{The role of the spin in quasiparticle interference}
\author{J. I. Pascual$^{1,2}$,
G. Bihlmayer$^5$,Yu. M. Koroteev$^{3,4}$,
 H.-P. Rust$^{2}$, G. Ceballos $^{2}$, M. Hansmann $^{2}$,  K. Horn $^{2}$,
E. V. Chulkov$^{3,6}$, S. Bl{\"u}gel$^5$, P. M. Echenique$^{3,6}$, and Ph.
Hofmann$^7$} 
\affiliation{
$^1$ Institut f\"ur Experimentalphysik, Freie Universit\"{a}t Berlin, 14195 Berlin, Germany\\
$^2$ Fritz-Haber-Institut der Max-Planck-Gesellschaft, 14195 Berlin, Germany\\
$^3$Donostia International Physics Center (DIPC), 20018 San Sebasti{\'a}n/Donostia, Basque Country, Spain\\
$^4$Institute of Strength Physics and Materials  Science, Russian
Academy of Sciences, 634021, Tomsk, Russia\\
$^5$Institut f\"ur Festk\"orperforschung, Forschungszentrum
J\"ulich, 52425 J\"ulich, Germany\\
$^6$Departamento de F\'{\i}sica de Materiales and Centro Mixto CSIC-UPV/EHU,
Facultad de Ciencias Qu\'{\i}micas, UPV/EHU, 20080 San Sebasti{\'a}n/Donostia,
Basque Country, Spain\\
$^7$Institute for Storage Ring Facilities, University of Aarhus,
8000 Aarhus C, Denmark}
\email[]{philip@phys.au.dk}
\homepage[]{http://www.phys.au.dk/~philip/}
\date{\today}
\begin{abstract}
Quasiparticle interference patterns measured by scanning tunneling microscopy (STM)
can be used to study the local electronic structure of metal surfaces and high
temperature superconductors. Here, we show that even in non-magnetic systems the
spin of the quasiparticles can have a profound effect on the interference patterns.
On Bi(110), where the surface state bands are not spin-degenerate, the patterns are
not related to the dispersion of the electronic states in a simple way. In fact, the features
which are expected for the spin-independent situation are absent and the
observed interference patterns can only be interpreted by taking spin-conserving scattering events into account.
\end{abstract}


\maketitle
The coupling of spin and orbital angular momenta leads to spin-dependent phenomena even in non-magnetic materials \cite{Winkler:2003}. For example, spin-orbit coupling (SOC) can remove the spin degeneracy of electronic bands and split them. This effect \cite{Rashba:1960} can be used to inject spin-polarized currents into spintronic devices from a non-magnetic quantum well structure \cite{Koga:2002}. On metal surfaces a strong SOC can profoundly change the Fermi surface, screening and electron dynamics \cite{Koroteev:2004}. Despite the variety of manifestations of the SOC effect in condensed matter, some fundamental questions remain unresolved. In particular, how does the spin affect the quasiparticle interference (QI)\cite{Petersen:2000c}? One possible approach to investigate the QI is the observation of oscillatory patterns in the local density of states with a scanning tunneling microscope (STM) \cite{Crommie:1993}. Especially, a Fourier transformation (FT) of STM conductance images can give valuable information about the electronic structure and Fermi surface of quasi two-dimensional systems
\cite{Sprunger:1997,Hofmann:1997,Petersen:1998,McElroy:2003}.
The analysis of QI patterns by STM has found a wide range of
applications such as probing the electronic structure of nano-scale objects
\cite{Diekhoener:2003}, contacting of molecular wires \cite{Moresco:2003} as
well as measuring the superconducting energy gap function and possible local
ordering on high-temperature superconductors \cite{Hoffman:2002,
McElroy:2003,Vershinin:2004}. However, only little attention has so far been payed to the fact that the quasiparticles can also have a spin.

Here, we show that the quasiparticle spin can have dramatic effects on the
interference patterns. In particular, it completely suppresses the interference
events associated with the Fermi surface which are usually observed on metal
surfaces\cite{Crommie:1993,Sprunger:1997,Hofmann:1997,Petersen:1998}. Instead, the patterns are dominated by spin-conserving processes. We use the (110)
surface of Bi to illustrate this.  Bi(110) is ideal for such experiments for two
reasons: First, it can be viewed as a two-dimensional metal because it supports
a number of surface states crossing the Fermi level ($E_\mathrm{F}$)
\cite{Agergaard:2001}, whereas the bulk density of states at $E_\mathrm{F}$ is
negligible for our purposes.  Second, the spin-orbit (SO) interaction leads to a
strong splitting of the surface state bands \cite{Koroteev:2004},
resulting in bands which are non-degenerate with respect to spin.

The Bi(110)
surface was cleaned by sputtering and annealing. STM data were taken using a low
temperature microscope \cite{Rust:1997}  at 5 K. 
The calculations have been performed
by using the full-potential linearized augmented planewave method in
film-geometry as implemented in the {\sc fleur} program
and local density approximation for the description of
exchange-correlation potential.
Technical details can be found in Ref. \cite{Koroteev:2004}.

\begin{figure}
\includegraphics[scale=1.2]{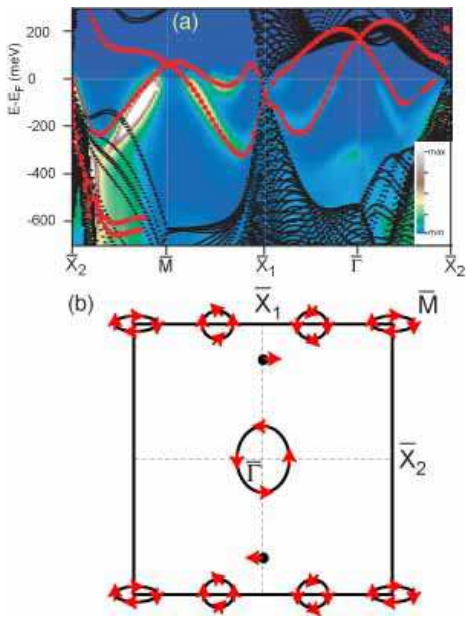}
  \caption{(a) Surface electronic structure of Bi(110). The color scale plot is the linear
  photoemission intensity measured by ARPES, published in Ref. \cite{Agergaard:2001},
  the red markers are the result of our {\em ab initio} calculation. (b) sketch of the SBZ and the Fermi surface
with an indication of the approximate spin-direction.
    \label{fig:1}}
\end{figure}

In order to lay the ground for an interference pattern analysis on Bi(110), consider its electronic structure as determined by
angle-resolved photoemission spectroscopy (ARPES) and first-principles calculations
(Fig.~\ref{fig:1}(a)).
The agreement between ARPES and calculations is excellent, but only if the  SO
interaction is taken into account in the latter. At the surface, the SO interaction lifts the
Kramers degeneracy and all the
 calculated bands are non-degenerate, i.e.\ they contain only
one spin per band and $k$ (momentum) point. Only at some special points of the surface
Brillouin zone (SBZ) such as $\bar{\Gamma}$ and $\bar{\mathrm{M}}$, symmetry forces
the two spin-split bands to be degenerate \cite{Koroteev:2004}. The large splitting
results in a Fermi surface (FS) with four distinct, non spin-degenerate elements
(see Fig.~\ref{fig:1}(b)). There are two hole pockets around the $\bar{\Gamma}$ and
$\bar{\mathrm{M}}$ points, a shallow electron pocket centered on the
$\bar{\mathrm{M}}$-$\bar{\mathrm{X}}_{1}$ line and a very small feature along
$\bar{\Gamma}$-$\bar{\mathrm{X}}_{1}$ which can be viewed as a single point for the
present purpose. The approximate spin directions on the Fermi surface are indicated
by red arrows. In the most simple picture, the spin is oriented in the plane of the
surface and perpendicular to the propagation direction. This is exactly true for a
two-dimensional gas of free electrons where the SO interaction can be treated by
adding a so-called Rashba-term to the Hamiltonian \cite{Rashba:1960}. Here the spin
directions can be worked out by inspecting the band dispersion in
Fig.~\ref{fig:1}(a). For example, the hole pocket states near $\bar{\Gamma}$ and
$\bar{\mathrm{M}}$ are formed by the lower branch of the SO split bands. Therefore,
the relative orientation between the two-dimensional momentum $\vec{k}$ and the spin
has to be the same for both states. Hence, the indicated direction of the spin has
to rotate anti-clockwise and clockwise around the $\bar{\Gamma}$ and
$\bar{\mathrm{M}}$ contours, respectively. For the electron pocket along
$\bar{\mathrm{M}}$-$\bar{\mathrm{X}}_{1}$ the spin polarization merely "wiggles"
around the preferred direction because of the need to take into account the
symmetry-requirement imposed by the crossing of the SBZ boundary.

In a simple picture, a quasiparticle interference pattern near $E_\mathrm{F}$
 arises because an electron
with wave vector $\vec{k}_\mathrm{F}$ encounters a defect such as an impurity or a
step edge and is reflected into a state with wave vector $-\vec{k}_\mathrm{F}$. The
interference of incoming and reflected waves gives rise to a modulation in the local
density of states with a periodicity of $2\vec{k}_\mathrm{F}$, i.e.\ with the vector
connecting the two states. A Fourier
transform of STM differential conductance ($dI/dV$) maps near $E_\mathrm{F}$, therefore,
shows an image of the two-dimensional Fermi surface with high intensities at points
with $2\vec{k}_\mathrm{F}$  \cite{Sprunger:1997,Hofmann:1997,Petersen:1998}. On high-temperature superconductors, similar
interference patterns are observed but the intensities in the FT maps are strongly
influenced by the density of the states causing the scattering events
\cite{McElroy:2003}. 

The spin map from Fig.~\ref{fig:1}(b) raises questions about
the possible quantum interference phenomena in the present case because
states with $\vec{k}_\mathrm{F}$ and $-\vec{k}_\mathrm{F}$ have opposite spin
directions. Therefore, all of the interference processes
building up the FS topology in the FT maps should be forbidden
\cite{Petersen:2000c}.

\begin{figure}
\includegraphics[scale=1.1]{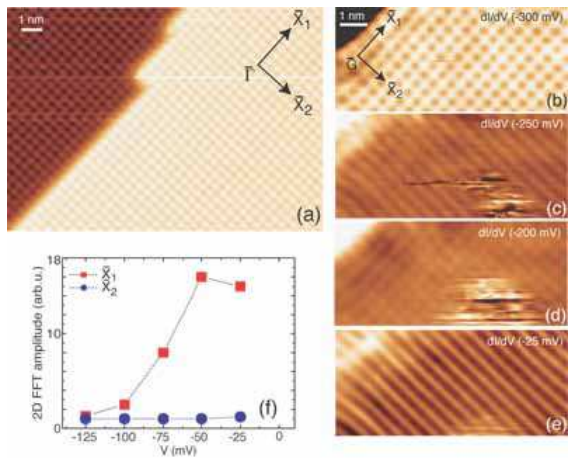}
  \caption{(a) Atomically-resolved constant current images of Bi(110) near a step ($V$=-205~mV;$ I=$~1~nA); (b)-(e) Conductance images taken at different bias voltages ( $I=$~1~nA); (f) Intensity of the
lattice spots in the FT conductance images as a function of bias voltage. 
 \label{fig:2}}
\end{figure}

In order to test this hypothesis,
we now analyze the interference patterns on Bi(110). 
Figs.~\ref{fig:2}(a) shows atomically-resolved images of the surface topography
close to a step on the surface. Bi(110) consists of a bilayer-like structure with an
almost square unit cell, containing two atoms \cite{Agergaard:2001}. One of the
atoms binds to the underlying layer; the other atom presents a  dangling bond. The
STM resolves only this latter atom.

The energy-dependent quasiparticle interference can be studied by $dI/dV$
 maps. Such maps are shown in Fig.~\ref{fig:2} (b)-(e) for
different bias voltages. For bias voltages below the onset of the surface state
bands ($V<-$300 mV), the maps show periodic structures very similar to the
topography images. Above the onset ($V>-$300 mV), weak sine-like modulations
along $\bar{\Gamma}$-$\bar{\mathrm{X}}_{1}$ and
$\bar{\Gamma}$-$\bar{\mathrm{X}}_{2}$ become clearly visible. Finally,  $dI/dV$ maps
close to $E_\mathrm{F}$ show a pronounced modulation along
$\bar{\Gamma}$-$\bar{\mathrm{X}}_{1}$. Fig.~\ref{fig:2} (f) compares the strength of
the lattice-periodic modulations as a function of the bias voltage close to $E_\mathrm{F}$. 
It displays the intensity of the corresponding
spots in the FT of a set of conductance images. It is evident that an enhancement
of the modulation only along the $\bar{\Gamma}$-$\bar{\mathrm{X}}_{1}$ direction sets in  for bias voltages above $-$100~meV.

It is important to realize that the lattice spots are a genuine spectroscopic feature and
not merely caused by the geometric 
structure of the surface.
Their origin is related to the interference of
Bloch wave-type surface states
\cite{Briner:1997b,Petersen:2000f,Pascual:2001}. In the present context, they have a
particular importance because some umklapp-type interference processes, involving a reciprocal lattice
vector $\vec{g}$, are not forbidden by
spin. For example, interference between $(\vec{k}_\mathrm{F},\leftarrow)$ and
$(\vec{k}_\mathrm{F}+\vec{g},\leftarrow)$ is always permitted while interference
between $(\vec{k}_\mathrm{F},\leftarrow)$ and $(-\vec{k}_\mathrm{F},\rightarrow)$ is
not. On the basis of this, it is easy to interpret the strong change of the
$\bar{\mathrm{X}}_{1}$ lattice spot intensity in Fig.~\ref{fig:2}(f) around 
$-$75~meV.  An inspection of the band structure in Fig.~\ref{fig:1}(a) reveals that the
bottom of the shallow electron-pocket along
$\bar{\mathrm{X}}_{1}$-$\bar{\mathrm{M}}$ happens to be at this energy. For lower energies, the electron pocket can not contribute to any interference events and the intensity of the $\bar{\mathrm{X}}_{1}$ lattice spots drops. 

\begin{figure}
\includegraphics[scale=0.5]{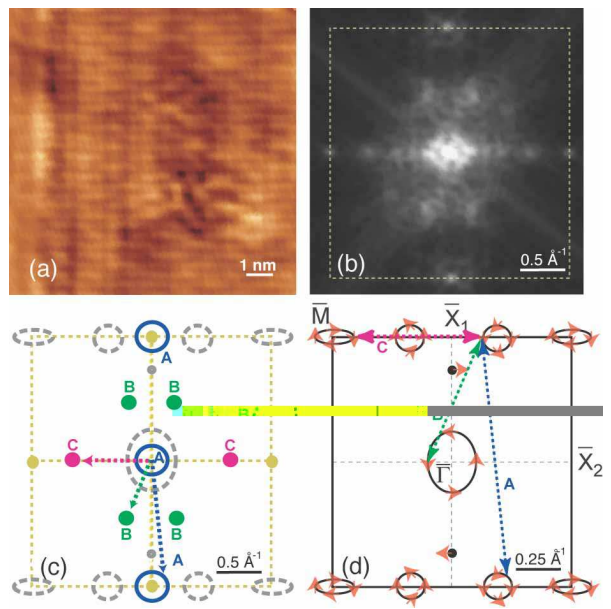}
  \caption{  (a) Differential conductance map ($V=$~40~mV;$ I=$~1~nA) and (b) its  FT map. (c) Schematic drawing of the FT map. The dashed grey lines are the modulations which would be expected for quasiparticle interference from a Fermi surface in Fig.~\ref{fig:1}(b) if the spin was not important. The features A,B and C are the actually observed non lattice-periodic structures. The yellow markers represent the lattice spots. (d) Illustration of the spin-conserving scattering events causing features A, B and C.
The dashed scattering vectors have to be translated to the origin and scaled down by a factor of 2 to yield the features in (c).
  \label{fig:3}}
\end{figure}

To study the QI away from the lattice spots,
figs.~\ref{fig:3}(a) and (b) show a typical $dI/dV$ image taken at $V$=+40 mV
together with its Fourier transformation. The bias voltage is sufficiently small to
allow a direct comparison of the observed interference features with the Fermi
surface sketched in Fig.~\ref{fig:1}(b). Fig.~\ref{fig:3}(c) shows a schematic
drawing of the FT map (colored features) together with the expected
interference pattern for a spin-independent situation (grey dashed
lines). It is evident that the latter is totally absent from the FT map, in
accordance with the simple picture outlined above. Instead, a number of other
features is observed in addition to the lattice spots.

 The first of these, labelled A in Fig.~\ref{fig:3}(c), is a fuzzy
ellipse around the origin $\bar{\Gamma}$, which can be also detected around the
$\bar{\mathrm{X}}_{1}$ points. The longer radius is of the order of 0.2 \AA$^{-1}$.
The second feature, labelled B in the figure, consists of four spots at a distance
of 0.62 \AA$^{-1}$ and 0.26 \AA$^{-1}$ from $\bar{\Gamma}$ in the $\bar{\Gamma}$-$\bar{\mathrm{X}}_{1}$ and $\bar{\Gamma}$-$\bar{\mathrm{X}}_{2}$ directions, respectively. Features C are two spots on either side of
the center, lying in the $\bar{\Gamma}$-$\bar{\mathrm{X}}_{2}$ direction. Their
distance from $\bar{\Gamma}$ is  0.9 \AA$^{-1}$.

We can explain these additional features by considering only the scattering
processes which conserve the spin. Some of these are indicated by
dashed lines in Fig.~\ref{fig:3}(d). They lead to interference patterns with a periodicity of 
$\vec{q}=\vec{k}_{\mathrm{final}}-\vec{k}_{\mathrm{initial}}$. The resulting $\vec{q}$~'s are also shown in Fig.~\ref{fig:3}(c), taking the scaling difference of (c) and (d) into account. Using these vectors, we can identify the 
B and C features as caused by
 interference between the electron pockets along the
$\bar{\mathrm{M}}$-$\bar{\mathrm{X}}_{1}$ direction and hole pocket states around
$\bar{\Gamma}$ and $\bar{\mathrm{M}}$, respectively. For the A structures the most
likely origin are scattering processes between states in the electron pockets along
$\bar{\mathrm{M}}$-$\bar{\mathrm{X}}_{1}$. Interference due to these scattering
processes would be expected to show up as elliptical features around the
$\bar{\mathrm{X}}_{1}$ lattice spots only. It is well known, however, that shifted
replicas of such features can also be found around the origin. These shifted
replicas can be much more intense than the originals around the higher order lattice
spots, depending on the character of the surface wave function
\cite{Petersen:2000f}.

Our findings show that taking the spin into account can be essential for the
understanding of QI.
Indeed, the interference patterns on Bi(110) cannot be directly related to the Fermi surface
topology as it happens for spin-degenerate bands \cite{Petersen:1998}, nor can they
be understood in terms of wave-vectors connecting points of a high density of states
\cite{Hoffman:2002, McElroy:2003}. The absence of
these "normal" interference patterns can be explained by taking the spin direction into account. Instead, we
find strong, energy-dependent lattice spots and spectroscopic signatures of 
spin-conserving interference patterns.

The formation of standing electron waves in a situation where the spin plays a role
was initially discussed for the surface states on Au(111).
 As was found by ARPES and confirmed by
first-principles calculations, this surface state band is split by the SO
interaction \cite{LaShell:1996,Nicolay:2001}. The splitting leads to a lift of the
spin degeneracy and changes the dispersion from a simple parabolic band to two
slightly shifted parabolas. The Fermi surface is turned from a circle into two
concentric circles of nearly the same radius. An analysis of QI close to the Fermi level, however, revealed only one circle \cite{Petersen:1999d} and this was
explained by the fact that quasiparticles with opposite spin cannot give rise to
observable interference patterns \cite{Petersen:2000c}. 
In this case, the spin seems to induce only a small error in the Fermi surface determined by QI. Our results show, however, that the observed QI pattern can be profoundly changed in the case of strong spin-orbit splitting and an interpretation of our data as an image of the Fermi surface would be completely misleading.

The consequences from the present work are twofold: First, it is obviously essential
to take the spin into account when interpreting quasiparticle interference. This
will be especially relevant for complex system with low symmetry where the spin
character of the states is not known \emph{a priori}. More importantly, our results
open entirely new possibilities by showing that the spin character of the
quasiparticles is reflected in the screening of defects on the surface and can be
probed by a non-magnetic STM. The combination of known Fermi surface and observed
interference pattern gives direct information about the relative spin of the
quasiparticles at the Fermi surface. Such information can be important, for example
when discussing different pairing mechanisms leading to superconductivity or the
possibility of spin-charge separation.

\begin{acknowledgments}
This work was partially supported by the UPV/EHU, Spanish MCyT and by the Danish National Science Foundation.
\end{acknowledgments}

\end{document}